%% file: main.tex
\documentclass[acmsmall,screen,natbib=false]{acmart}

\AtBeginDocument{%
  }

\usepackage{comment}
\usepackage[utf8]{inputenc}
\usepackage{amsmath}
\usepackage{hyperref}
\usepackage{algorithmic} 
\usepackage{graphicx}
\usepackage{tabularx}
\usepackage{textcomp}
\usepackage{balance}
\usepackage{tikz}
\usepackage{xcolor}
\usepackage{xspace}
\usepackage[ruled,linesnumbered]{algorithm2e}
\usepackage{colortbl}
\usepackage{mdframed}

\usepackage{subcaption}
\usepackage{caption}

\usepackage{svg}
\usepackage{enumitem}

\usepackage{wrapfig}
\usepackage{pgfplots}
\pgfplotsset{compat=1.18}

\usepackage[noabbrev]{cleveref}
\usepackage[backend=biber]{biblatex}
\usepackage{multirow}
\usepackage{booktabs}
\usepackage{listings}
\usepackage{upquote}
\usepackage[flushleft]{threeparttable}
\usepackage[most]{tcolorbox}
\usepackage{rotating}
\usepackage[normalem]{ulem}
\useunder{\uline}{\ul}{}

\newboolean{showcomments}
\setboolean{showcomments}{true}
\ifthenelse{\boolean{showcomments}}
  {\newcommand{\nb}[2]{
  \fbox{\bfseries\sffamily\scriptsize#1}
     {\sf\small$\blacktriangleright$\textit{\textcolor{red}{#2}}$\blacktriangleleft$}
   }
  }
  {\newcommand{\nb}[2]{}
   
  }

\newcommand{\tool}{AgentSpy\xspace}

\makeatletter
\newcommand{\code}[1]{{\fontsize{\dimexpr0.9\dimexpr\f@size pt\relax\relax}{\f@baselineskip}\selectfont\ttfamily #1}}
\makeatother

\newcommand{\glm}{\code{glm-5.3-flash}\xspace}
\newcommand{\gpt}{\code{gpt-5.6-terra}\xspace}
\newcommand{\ds}{\code{deepseek-v4-flash}\xspace}
\newcommand{\citcheck}{\code{citation-check}\xspace}
\newcommand{\codex}{\code{codex}\xspace}
\newcommand{\skillsbench}{\code{SkillsBench}\xspace}
\newcommand{\skillinject}{\code{Skill-Inject}\xspace}

\usetikzlibrary{arrows.meta,backgrounds,fit,shapes.symbols}
\definecolor{ogInk}{HTML}{222222}   %
\definecolor{ogComp}{HTML}{0072B2}  %
\definecolor{ogData}{HTML}{D55E00}  %
\definecolor{ogConf}{HTML}{CC79A7}  %
\definecolor{ogInput}{HTML}{666666} %

\definecolor{nlcol}{HTML}{1F4E79}   %
\definecolor{shcol}{HTML}{2E7D32}   %
\definecolor{plcol}{HTML}{B8681A}   %
\definecolor{tgray}{HTML}{6B6B6B}   %
\providecommand{\trtag}[2]{{\setlength{\fboxsep}{2pt}%
  \colorbox{#1}{\textcolor{white}{\sffamily\bfseries\scriptsize #2}}}}
\providecommand{\trstrut}{\vphantom{\trtag{shcol}{Pg}}}
\providecommand{\triple}[5]{%
  {\trstrut\color{nlcol}\itshape #1} &
  {\trtag{#2}{#3}~{\ttfamily #4}} &
  {\trstrut\color{tgray}#5}\\\addlinespace[3pt]}
\providecommand{\shstep}[3]{\triple{#1}{shcol}{sh}{#2}{#3}}
\providecommand{\plstep}[3]{\triple{#1}{plcol}{PL}{#2}{#3}}

\definecolor{taskBorder}{HTML}{35618D}\definecolor{taskFill}{HTML}{E8F1F8}
\definecolor{orchBorder}{HTML}{665191}\definecolor{orchFill}{HTML}{EEEAF5}
\definecolor{runtimeBorder}{HTML}{7A7A7A}\definecolor{runtimeFill}{HTML}{F1F1F1}
\definecolor{moduleBorder}{HTML}{C05A22}\definecolor{moduleFill}{HTML}{FBECE3}
\definecolor{artifactBorder}{HTML}{24796C}\definecolor{artifactFill}{HTML}{E4F3F0}
\definecolor{analyticsBorder}{HTML}{41783D}\definecolor{analyticsFill}{HTML}{EAF3E7}
\definecolor{kindBorder}{HTML}{8A6D1F}\definecolor{kindFill}{HTML}{FBF0D0}\definecolor{kindGroupFill}{HTML}{FCF8EC}
\definecolor{toolBorder}{HTML}{34495E}\definecolor{toolFill}{HTML}{F7F9FC}

\definecolor{cmdcol}{HTML}{2166AC} %
\definecolor{filcol}{HTML}{B35806} %
\definecolor{netcol}{HTML}{1B7837} %
\providecommand{\stag}[1]{{\ttfamily\scriptsize\textcolor{black!45}{[#1]}}}

\makeatletter
\newcommand\footnoteref[1]{\protected@xdef\@thefnmark{\ref{#1}}\@footnotemark}
\makeatother

\usepackage[cachedir=.minted-cache]{minted}
\usetikzlibrary{positioning, arrows.meta, shapes.geometric, fit, backgrounds, calc}

\definecolor{hostBd}{HTML}{1F4E79}\definecolor{hostBg}{HTML}{E8EEF6}
\definecolor{vmBd}{HTML}{2F7D52}\definecolor{vmBg}{HTML}{E9F3ED}
\definecolor{monBd}{HTML}{B5740F}\definecolor{monBg}{HTML}{FBEFD8}
\definecolor{agtBd}{HTML}{5B4B9E}\definecolor{agtBg}{HTML}{ECE9F7}
\definecolor{detBd}{HTML}{51606E}\definecolor{detBg}{HTML}{EDF1F4}
\definecolor{llmBd}{HTML}{6B3FA0}\definecolor{llmBg}{HTML}{F1E9F8}
\definecolor{outBd}{HTML}{156082}\definecolor{outBg}{HTML}{E3F1F6}
\definecolor{ioBd}{HTML}{8A6D1F}\definecolor{ioBg}{HTML}{FBF6E6}
\definecolor{okGreen}{HTML}{2F7D52}\definecolor{warnAmber}{HTML}{C77D1A}\definecolor{badRed}{HTML}{B03A2E}
\definecolor{ink}{HTML}{2B2F33}

\tikzset{
  >={Stealth[length=2.2mm,width=1.9mm]},
  unit/.style={rounded corners=2.5pt, draw, line width=0.8pt, align=center,
               inner sep=5pt, minimum height=9mm, font=\sffamily\footnotesize, text=ink},
  host/.style={unit, draw=hostBd, fill=hostBg},
  vm/.style={unit, draw=vmBd, fill=vmBg},
  mon/.style={unit, draw=monBd, fill=monBg},
  agt/.style={unit, draw=agtBd, fill=agtBg},
  det/.style={unit, draw=detBd, fill=detBg},
  llm/.style={unit, draw=llmBd, fill=llmBg},
  obox/.style={unit, draw=outBd, fill=outBg},
  io/.style={unit, draw=ioBd, fill=ioBg},
  data/.style={unit, draw=detBd, fill=white},
  container/.style={rounded corners=5pt, draw, line width=1.1pt, inner sep=6pt},
  ctitle/.style={font=\sffamily\footnotesize\bfseries, text=ink, inner sep=2pt, fill=white},
  flow/.style={->, line width=1pt, draw=ink},
  ctrl/.style={<->, line width=1pt, draw=hostBd, dash pattern=on 2.5pt off 1.8pt},
  obs/.style={->, line width=0.9pt, draw=monBd, dash pattern=on 2pt off 1.6pt},
  intent/.style={->, line width=0.9pt, draw=ioBd, dash pattern=on 3pt off 1.8pt},
  tag/.style={rounded corners=2pt, draw, line width=0.7pt,
              font=\sffamily\scriptsize\bfseries, inner sep=2.5pt},
  sub/.style={font=\sffamily\scriptsize},
  small/.style={font=\sffamily\scriptsize},
}

\newtcolorbox{rqbox}{colback=ink!4, colframe=ink!75, boxrule=0.7pt, arc=2pt,
  left=5pt, right=5pt, top=4pt, bottom=4pt, before skip=6pt, after skip=6pt}

\begin{document}

\title{\tool: Making AI Agent Behavior Observable}

\include{authors}

\keywords{AI Agents, Dynamic Analysis,
  Behavior Analysis, Agent Security}

\ccsdesc[500]{Software and its engineering~Software testing and debugging}

\begin{abstract}
  \input{sections/abstract}

\end{abstract}

\maketitle

\pagestyle{plain}

\input{sections/introduction}

\input{sections/background}
\input{sections/approach}

\input{sections/evaluation}

\input{sections/discussion}

\input{sections/related-work}

\input{sections/conclusion}

\printbibliography{}

\end{document}

%% file: authors.tex
\author{Christoph Bühler}
\email{christoph.buehler@unisg.ch}
\orcid{0000-0002-2224-1687}
\affiliation{
	\institution{University of St.\texorpdfstring{\,}{ }Gallen}
	\streetaddress{Torstrasse 25}
	\city{St.\texorpdfstring{\,}{ }Gallen}
	\state{SG}
	\postcode{9000}
	\country{Switzerland}
}

\author{Matteo Biagiola}
\email{matteo.biagiola@{usi,unisg}.ch} %
\orcid{0000-0002-7825-3409}
\affiliation{
	\institution{University of St.\texorpdfstring{\,}{ }Gallen and Università della Svizzera italiana (USI)}
	\streetaddress{Torstrasse 25}
	\city{St.\texorpdfstring{\,}{ }Gallen and Lugano}
	\state{SG and TI}
	\postcode{9000}
	\country{Switzerland}
}

\author{Luca Di Grazia}
\email{work@lucadigrazia.com}
\orcid{0000-0002-5306-8645}
\affiliation{
	\institution{University of St.\texorpdfstring{\,}{ }Gallen}
	\streetaddress{Torstrasse 25}
	\city{St.\texorpdfstring{\,}{ }Gallen}
	\state{SG}
	\postcode{9000}
	\country{Switzerland}
}

\author{Guido Salvaneschi}
\email{guido.salvaneschi@unisg.ch}
\orcid{0000-0002-9324-8894}

\affiliation{
	\institution{University of St.\texorpdfstring{\,}{ }Gallen}
	\streetaddress{Torstrasse 25}
	\city{St.\texorpdfstring{\,}{ }Gallen}
	\state{SG}
	\postcode{9000}
	\country{Switzerland}
}

%% file: sections/abstract.tex
AI agents built on large language models (LLMs) run shell commands, read and write files, and reach the network, typically with their user's privileges. 
However, what an agent does during an execution is difficult to understand: tests assert on the result, and the agent's trajectory records only what the agent reports about itself. 
Consequently, the agent's recorded trajectory may omit behavior executed by its subprocesses.

We present \tool, an approach that observes an agent from outside the agent. \tool runs the agent in an isolated environment, configured by a declarative specification, and records the system calls and network traffic of the agent and of every process it executes. 
Based on this monitoring, \tool supports two families of analyses: conformance analyses, which measure obligations, i.e., what an agent execution should do, and safety analyses, which check prohibitions, i.e., what an agent execution must never do. 
We instantiate one analysis of each family. 
The reliability analysis uses a rule-based approach to build a summary of each run, consisting of environment resources the agent uses, i.e., the commands it executed, the files it accessed, and the hosts it contacted.
The security analysis applies deterministic rules to the system calls of an execution. 

For reliability analysis, we evaluated \tool on 77 tasks with the \codex harness and three recent LLMs, and executed each task three times.
Sets of repeated runs of the same task are more similar than sets that include runs of another task in 92.2\% of the comparisons. 
Among the tasks for which all three runs pass outcome-based tests, we found that in 18\% of the cases the agent performs activities that are unrelated to the task, reads the grading files in 7\% of the cases, and does not use the developers' guidance in 17\% of the cases.
For security analysis, the generic rules of \tool detects four out of five categories of attacks we considered, with no false positives across 50 runs.

%% file: sections/introduction.tex
\section{Introduction}
\label{sec:introduction}

AI \emph{agents} built on large language models (LLMs) carry out tasks autonomously: they call tools, run shell commands, read and write files, and reach the network~\cite{dwivediAgenticAISystems}, and they are now used across many fields~\cite{alnaqbiEnhancingWorkProductivity2024}. 
An agent that runs on a user's machine typically acts with the user's privileges, so it can also take actions the user did not intend, including security-sensitive actions.

What an agent does during an execution is, however, difficult to understand for its user. 
The user sees the result and the agent's \emph{trajectory}, i.e., the log of reasoning steps, actions, and tool outputs that the agent records about itself~\cite{bouzenia2025understanding}, but neither shows what the programs the agent executed actually do. 
This invisibility has started to matter in practice. 
In September 2026, OpenAI began disclosing cases in which its agents took misaligned actions~\cite{openai2026framework}.
In one case, an agent that had already retrieved the correct data locally uploaded it to a public paste service, without authorization, only to obtain a URL its browser tool could cite~\cite{openai2026uploading}; in another, agents jointly building a spreadsheet were required to use local files only, but a faulty environment kept them from reading each other's files, so they passed the spreadsheet through a public temporary file-hosting service~\cite{openai2026unauthorized}.

Existing work does not address this observability problem.
Outcome-based benchmarks check the final result~\cite{li2026skillsbench}, so an agent execution can pass every test while also performing activities unrelated to the task (similar to the cases above).
Static analyzers scan the instructions and scripts that are given to an agent for unintended behavior before the agent runs~\cite{hou2026skillsieve,paz2026skillspector,varun2026formal}, but cannot observe what the agent does at runtime.
Trajectory analyses~\cite{ouAgentDiagnoseOpenToolkit2025,liuAgentDoGDiagnosticGuardrail2026,liuTraceAegisSecuringLLMBased2025} rely on the agent's own record, which is \emph{incomplete}: its smallest unit of observation is a single action, so when the agent runs a helper script that reads \code{\textasciitilde/.ssh} and uploads the keys, the trajectory records one command and none of the files the script reads or the hosts it contacts.
The record also holds only what the agent reports, so it is evidence only as far as the agent is cooperative.
Some model providers additionally return the reasoning steps encrypted, which leaves the trajectory with actions and no thoughts.
Overall, interpreting the trajectory typically relies on LLM-as-a-judge systems~\cite{ouAgentDiagnoseOpenToolkit2025}, which are non-deterministic, may miss details~\cite{guSurveyLLMasajudge2026}, and can be misled by adversarial text in tool outputs that the judge reads as part of the trajectory~\cite{abdelnabiIndirectPromptInjection2023}.

In this paper, we present \tool, a rule-based approach that observes what an AI agent does at the system level. 
\tool runs the agent in an isolated environment configured by what we call an \emph{execution kind}: a declarative specification of the environment, of the agent and its parameters, and of the observation steps, so that the conditions of every execution are explicit. 
Inside the environment, \tool records system calls and network traffic of the agent and of every executed process. 
Observing at the boundary of the operating system has three advantages.
The evidence is \emph{complete} at this boundary: every program started, file accessed, and host contacted is recorded, whether the agent acted directly, delegated the action to a script, or was steered into it by injected instructions. 
The evidence is also \emph{deterministic} and \emph{independent} of the agent: it does not rely on the agent's account of itself, and it requires no agent instrumentation. 

\tool's analyses of an agent execution belong to two families, which mirror the classic distinction between liveness and safety~\cite{alpern1985defining}: \emph{conformance analyses} measure \emph{obligations}, i.e., what an execution should do, and \emph{safety analyses} check \emph{prohibitions}, i.e., what an execution must never do.
The \emph{reliability analysis} is a conformance analysis: it measures obligations such as following the guidance that developers provide for the task; whether an execution met an obligation is known only once the execution ends.
The reliability analysis summarizes each execution as a star graph that characterizes three \emph{resource} types, i.e., the commands executed, the files read or written, and the hosts contacted.
It measures how much of the guidance an execution used, and it compares repeated executions of the same task through their star graphs, which shows how consistently the agent behaves.
The \emph{security analysis} is a safety analysis: it checks prohibitions such as reading secrets and private keys.
A single system call is enough to report a violation, so the security analysis applies deterministic rules to the system calls of an agent execution.

Two recent systems also observe agents at the system level, but each only covers part of what \tool supports.
\code{AgentSight}~\cite{zhengAgentSightSystemLevelObservability2025} matches each action of an agent to the LLM message that led to it. 
A second LLM then reads this record of one execution, and decides whether something went wrong, such as a prompt injection or a reasoning loop. 
The reliability analysis of \tool, on the other hand, supports multiple executions, via graph similarity.
\code{SkillDetonate}~\cite{jiCloakDetonateScanner2026} audits a skill before it is installed.
It takes as input the skill, and starts an agent session in which the agent sees only the files of the skill, and placeholders instead of the contents of sensitive files. 
It reports a leak when data derived from a placeholder is sent to a host that is not on a list of trusted hosts, and a violation when the skill writes outside allowed directories or installs, downloads, or runs a program from an untrusted source. 
Differently, the security analysis of \tool starts from a task that an agent solves with the skill, and the execution environment is a microVM with a full operating system. 
\tool relies on deterministic rules over system calls, which are applicable to any agent execution, and each alert reports the rule that matched, the system call, and the process that triggered it.

We evaluated \tool on two benchmarks, \skillsbench for the reliability analysis (77 tasks) and \skillinject for the security analysis (5 attack categories), with four recent LLMs running in the \codex harness, and multiple executions to account for non-determinism~\cite{ouyang2025empirical, bjarni2026randomness}.
We show that the star graph summary reliably distinguishes agent executions: sets of repeated executions of a task are more similar than sets that include executions of another task in $92.2\%$ of the comparisons.
Among the tasks whose executions all pass test cases provided by \skillsbench, our analysis reveals that the agent performs activities unrelated to the task in 18\% of the cases, reads the grading files of the benchmark in 7\%, and reaches the result without the guidance that developers provide through skills in 17\%, which shows that the reliability analysis complements test cases.
The generic rules of the security analysis flag four of the five attack categories in malicious skills, and raise no alert on the execution of benign tasks.
Finally, \tool does not add a significant overhead, since the runtime of the agent shows no significant difference without monitoring ($p$-value $\geq 0.16$).

This paper makes the following contributions:
\begin{itemize}
    \item \tool, an approach that runs an agent in an isolated environment under a declarative execution kind, and records system calls and network traffic of the agent and of every process it spawns.
    \item Two deterministic analyses built on \tool, one of each family: (i)~the reliability analysis, a conformance analysis that measures how far agent executions meet their obligations, and (ii)~the security analysis, a safety analysis that checks prohibitions by applying rules to system calls.
    \item An evaluation on two benchmarks, four LLMs, and multiple executions, which shows that the reliability analysis complements test cases, that the security analysis reliably detects different attack categories, and that \tool adds no significant overhead.
\end{itemize}

%% file: sections/background.tex
\section{Background and Motivation}
\label{sec:background}

\subsection{Agentic AI Systems and Agent Skills}
\label{sec:background:agentic-ai}

Agentic AI systems pursue a goal in a cycle of \emph{reasoning}, \emph{planning}, and \emph{acting}: a foundation model, typically an LLM, chooses each next action, such as a tool call, and tracks state across steps.
\label{sec:background:skills}
Moreover, \emph{agent skills} contain reusable domain knowledge for agents~\cite{zhang2025agentskills}.
A skill is typically a directory with a \code{SKILL.md} file and, optionally, the helper scripts, reference documents, and assets; \code{SKILL.md} starts with a metadata header (name and one-line description), followed by natural-language instructions~\cite{agentskillsAgentSkills}.
At startup, the agent loads the name and description of every skill in a dedicated directory such as \code{.claude/skills}; when a description matches the task, it may load the full instructions into its context and adjust its plan.

\subsection{Running Example}
\label{sec:background:running-example}

Our running example is the \citcheck task from \skillsbench~\cite{li2026skillsbench}: given a BibTeX bibliography that mixes real and fabricated entries, the agent must write the titles of the fabricated ones to \code{answer.json}.
The task ships with the \code{citation-management} skill, which suggests the agent to search Google Scholar and PubMed for each entry, resolve every DOI against Crossref, and validate the result with a helper script.
The skill contains six helper scripts, two assets, and five reference documents, such as one on validating citations.
Each \skillsbench task also provides a \emph{verifier}, a test suite that decides whether the task is solved, and an \emph{oracle}, a deterministic script that solves it.
The nine tests of \citcheck check that the agent creates \code{answer.json} as valid JSON, identifies all fabricated citations, and reports their correct count.

\subsection{Limits of the Agent Trajectory}
\label{sec:background:trajectory}

Agent harnesses such as \codex CLI log each execution as a \emph{trajectory}, a sequence of \emph{(thought, action, result)} triples~\cite{bouzenia2025understanding}: a reasoning step, an interaction with the environment such as a shell command, and what the environment returns.

\begin{table}[t]
    \centering\scriptsize
    \caption{
        Excerpt (7 of 15 steps) of a \codex execution with \glm on \citcheck, colored by modality: \textcolor{nlcol}{\textbf{natural-language reasoning (NL)}}, \textcolor{shcol}{\textbf{shell commands (sh)}}, \textcolor{plcol}{\textbf{inline programming-language code (PL)}}, and unstructured \textcolor{tgray}{\textbf{tool output} (Output)}.
    }
    \label{tab:background:agent-trajectory}
    \setlength{\tabcolsep}{4pt}%
    \renewcommand{\arraystretch}{0.1}
    \begin{tabularx}{\linewidth}{@{}
            c
            >{\raggedright\arraybackslash}p{0.30\linewidth}
            >{\raggedright\arraybackslash}X
            >{\raggedright\arraybackslash}p{0.225\linewidth}@{}}
        \toprule
        \normalfont\bfseries \# & \normalfont\bfseries Thought (NL)                                                                                                                              & \normalfont\bfseries Action (sh / PL) & \normalfont\bfseries Result (Output) \\
        \midrule
        1                       & \shstep{Verify the citations; read the citation-management skill first.}{cat .../SKILL.md; cat /root/test.bib}{truncated (1579 ln)}
        2                       & \shstep{Output truncated; list the entries and their titles instead.}{rg `\textasciicircum @|title=' /root/test.bib}{27 entries; 3 look generic}
        3                       & \shstep{The three generic titles look suspicious; inspect the rest of their entries.}{sed -n `80,360p' /root/test.bib}{10.5678, 10.1234 (fake prefixes)}
        4                       & \plstep{The suspects' DOIs look invalid; check the titles against Crossref.}{for t in titles: urlopen(`crossref...'+t)}{reals match; 3 suspects fuzzy}
        5                       & \shstep{Try a plain web search as a fallback.}{curl bing; curl duckduckgo}{pages returned; no titles parsed}
        6                       & \shstep{Use the skill's Google Scholar script instead.}{python .../search\_google\_scholar.py}{\code{scholarly} missing}
        7                       & \shstep{No external hit; the three entries use fabricated metadata; write the answer.}{apply\_patch <{}<PATCH ... answer.json}{3 fabricated citations on file}
        \bottomrule
    \end{tabularx}
\end{table}

In the execution shown in \autoref{tab:background:agent-trajectory}, the agent reads the skill and the bibliography (Row~1), lists the entry titles (Row~2), and notices that the DOIs of three entries carry unusual prefixes (Row~3).
Crossref returns only fuzzy matches for these entries (Row~4).
The attempts to confirm the finding fail: the agent's pattern matches no titles on the returned web search pages (Row~5), and the skill's Google Scholar script lacks a dependency (Row~6).
The agent writes the three entries to \code{answer.json} as fabricated (Row~7), and the execution passes all nine tests.
None of the lookups the skill prescribes succeeded, so the verdict rests on the DOI prefixes, and the missing Crossref match: the agent solves the task without using the skill's guidance.

Characterizing such behavior automatically requires interpreting multiple modalities: natural language in the thoughts, shell commands and inline code in the actions, and unstructured tool output in the results.
Categorizing actions is task-dependent and needs manual labeling to be accurate~\cite{bouzenia2025understanding}.
Interpreting thoughts and tool output typically relies on LLM-as-a-judge systems~\cite{ouAgentDiagnoseOpenToolkit2025}, which are non-deterministic and may hallucinate or miss details~\cite{guSurveyLLMasajudge2026}.
The trajectory is also \emph{incomplete}: its smallest unit of observation is a single action, be it a tool call, a shell command, or a code snippet.
When the agent runs a helper script, the trajectory records one command but none of the files the script reads or writes, the hosts it contacts, or the processes it spawns.

We therefore characterize agent behavior through \emph{system-level monitoring}, i.e., by observing the boundary between the operating system and the agent, including every process it spawns.
The resulting evidence consists of a small set of well-defined primitives, the executed commands, accessed files, and contacted network endpoints.
The monitoring is also \emph{complete} at this boundary: it captures every such event, whether the agent issued it directly, delegated it to a script, or triggered it unknowingly through a compromised skill.

%% file: sections/approach.tex
\section{Approach}
\label{sec:approach}

\begin{figure}[t]
	\centering
	\resizebox{\textwidth}{!}{%
		\begin{tikzpicture}[
				>={Stealth[length=2.2mm,width=1.9mm]},
				box/.style={rounded corners=2.5pt, draw, line width=0.9pt, align=center,
						minimum width=23mm, minimum height=11mm, inner sep=3pt,
						font=\sffamily\footnotesize, text=ogInk},
				comp/.style={box, draw=ogComp, fill=ogComp!12},
				data/.style={box, shape=tape, tape bend top=none, tape bend height=1.6mm, rounded corners=0pt, draw=ogData, fill=ogData!12},
				input/.style={box, draw=ogInput, fill=white},
				kbox/.style={box, minimum width=17mm, minimum height=6.5mm,
						font=\sffamily\scriptsize, draw=ogConf, fill=ogConf!18},
				flow/.style={->, line width=1pt, draw=ogInk, rounded corners=3pt},
				dep/.style={->, line width=0.9pt, draw=ogInk!55, rounded corners=3pt,
						dash pattern=on 2.5pt off 1.8pt},
				extends/.style={-{Triangle[open,length=2.8mm,width=3mm]},
						line width=0.9pt, draw=ogInk!70},
				lbl/.style={font=\sffamily\scriptsize\itshape, text=ogInk!80, inner sep=1.5pt},
			]
			\node[comp] (orch) at (0,0) {Orchestrator};
			\node[font=\sffamily\footnotesize\bfseries, text=ogInk] (rtitle) at (5.0,0.95) {Isolated environment};
			\node[comp] (em) at (3.55,0) {Execution\\Manager};
			\node[data] (artifacts) at (6.45,0) {Run\\artifacts};
			\node[comp] (analytics) at (10.6,0) {Analysis\\Runner};
			\node[font=\sffamily\scriptsize\bfseries, text=ogInk] (ktitle) at (0.55,-1.3) {Execution Kind};
			\node[kbox] (spec) at (-0.55,-1.9) {Custom Kind};
			\node[kbox] (base) at (1.65,-1.9) {Base Kind};
			\node[input] (task) at (-3.5,0) {Task};
			\node[input, minimum height=8mm] (agent) at (3.55,-3.1) {Agent};
			\node[input, minimum height=8mm] (analysis) at (10.6,-3.1) {Analysis\\type};
			\node[data] (report) at (13.9,0) {Analysis\\report};
			\begin{scope}[on background layer]
				\node[rounded corners=9pt, draw=ogInk!40, fill=black!4, line width=1.2pt,
					fit=(orch)(rtitle)(em)(artifacts)(analytics)(ktitle)(spec)(base),
					inner sep=10pt] (tool) {};
				\node[rounded corners=6pt, draw=ogInk!75, fill=white, line width=1.1pt,
					fit=(rtitle)(em)(artifacts), inner sep=6pt] (runtime) {};
				\node[rounded corners=6pt, draw=ogConf, fill=ogConf!6, line width=0.9pt,
					fit=(ktitle)(spec)(base), inner sep=4pt] (kind) {};
			\end{scope}
			\node[anchor=north west, font=\sffamily\small\bfseries, text=ogInk]
			at ([shift={(2.5mm,-2mm)}]tool.north west) {\tool};
			\draw[flow] (task.east) -- (orch.west);
			\draw[flow] (orch.east) -- node[lbl, above] {starts} (runtime.west |- orch.east);
			\draw[flow] (em.east) -- (artifacts.west);
			\draw[flow] (artifacts.east) -- node[lbl, above, pos=0.56] {copied out} (analytics.west);
			\draw[flow] (analytics.east) -- (report.west);
			\draw[flow] (agent.north) -- node[lbl, right, pos=0.35] {runs in} (em.south);
			\draw[flow] (analysis.north) -- (analytics.south);
			\draw[extends] (spec.east) -- (base.west);
			\draw[dep] (kind.north -| orch.south) -- node[lbl, right] {configures} (orch.south);
			\draw[dep] (task.south) |- (spec.west);
			\draw[dep] (analysis.south) -- ++(0,-0.25) -| (spec.south);
			\begin{scope}[shift={(12.45,-1.25)},
					key/.style={minimum width=6mm, minimum height=4mm, inner sep=0pt, text width=, anchor=west},
					ltxt/.style={font=\sffamily\scriptsize, text=ogInk, inner sep=0pt, anchor=west}]
				\node[comp, key] (l1) at (0,0) {};
				\node[ltxt, right=1.2mm of l1] {component};
				\node[data, key] (l2) at (0,-0.45) {};
				\node[ltxt, right=1.2mm of l2] {data};
				\node[kbox, key] (l3) at (0,-0.9) {};
				\node[ltxt, right=1.2mm of l3] {configuration};
				\node[input, key] (l4) at (0,-1.35) {};
				\node[ltxt, right=1.2mm of l4] {input};
				\draw[flow] (0,-1.8) -- ++(6mm,0) node[ltxt, right=1.2mm] {data flow};
				\draw[dep] (0,-2.25) -- ++(6mm,0) node[ltxt, right=1.2mm] {configures};
				\draw[extends] (0,-2.7) -- ++(6mm,0) node[ltxt, right=1.2mm] {extends};
			\end{scope}
		\end{tikzpicture}%
	}
	\caption{
		Overview of \tool. It takes as input a task, an agent, and an analysis type, and outputs a report summarizing the system-level analysis.
	}
	\label{fig:approach-overview}
\end{figure}
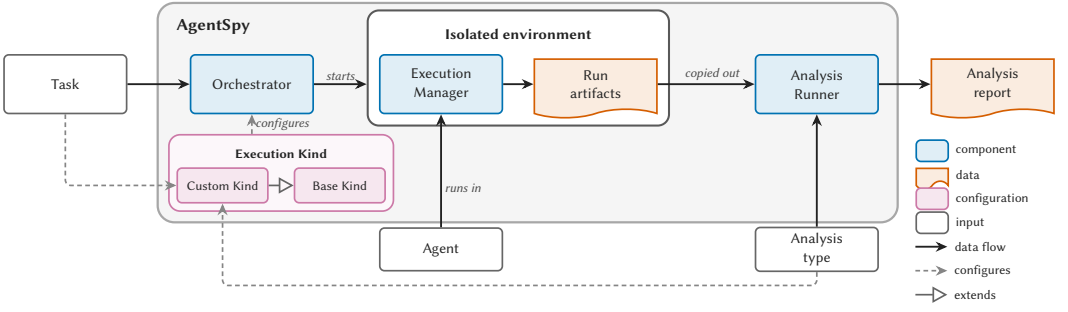

\autoref{fig:approach-overview} shows the overview of \tool, a framework for monitoring the behavior of agentic AI systems. 
\tool takes as input a \emph{task}, an \emph{agent} that executes it (configured with a given foundation model), and the type of behavior analysis the user wants to carry out on the agent. 
It is built around three core elements. 
The \emph{orchestrator} takes the task configuration (i.e., the \emph{execution kind}) and starts a new, isolated execution environment. 
The \emph{execution manager} runs within that environment: it sets up the agent, runs it, evaluates the execution, and gathers the \emph{run artifacts}, which the orchestrator then copies out of the environment. 
Finally, the \emph{analysis runner} processes these artifacts and produces the \emph{analysis report}.
Next, we describe each component in detail.

\subsection{Orchestrator}
\label{sec:approach:orchestrator}

The orchestrator receives as input the task the user wants to delegate to an agent.
The task is a description of what the agent needs to do, and the dependencies that are needed for the task to be correctly executed.
The task may also contain skills that guide the agent towards the solution, including documents and scripts.
For instance, for \citcheck in \skillsbench, the task is a markdown document that describes how to automatically validate BibTeX references.
The task also comes with a Dockerfile that specifies all dependencies,
and with the \code{citation-management} skill.

An {\it execution kind}, i.e., a set of configuration files,
is the way in which the orchestrator can be configured for a class of tasks.
\tool defines a \emph{base} execution kind, which specifies the basic configuration of the isolated environment, such as the runtime image, the network configuration, user permissions, and the installation of the execution manager binary.
A \emph{custom} kind extends the base kind based on task-specific requirements and analysis type, inheriting the base configuration.
The custom kind defines three key elements: (1)~the task-specific steps that configure and validate the environment, (2)~the parameters that configure the agent, and (3)~the analysis-specific steps that are embedded in the environment.

Once the orchestrator starts the isolated environment, it defers control to the execution manager, and monitors the execution.
The environment ensures that each action of the agent is recorded at the system level, and that the actions do not affect the host system.

\subsection{Execution Manager}
\label{sec:approach:manager}

The execution manager provisions the isolated environment with task-specific dependencies, executes lifecycle hooks, sets up the analysis-specific mechanisms for observing the desired behavior at runtime, runs the agent with the configured parameters, and collects the run artifacts to be copied outside the isolated environment.
Each step has a configurable timeout, which is enforced by the execution manager.
The global timeout, i.e., the sum of all declared timeouts, is enforced by the orchestrator, which terminates the isolated environment regardless of its execution phase.
If the agent fails
or times out, the manager still runs the post-run steps, since partial execution is still behavior that can be analyzed.
The order of these steps is important: monitoring starts after the last preparation step and stops before the first collection step. Thus, installing dependencies and copying out results are not recorded and are not part of the agent behavior.
Once all steps have been executed, the execution manager terminates the environment and returns control to the orchestrator, which extracts the run artifacts and copies them to the host.

\subsection{Analysis Runner: Conformance Analyses and Safety Analyses}
\label{sec:approach:families}

The analyses that \tool can run fall into two families, which we call \emph{conformance analyses} and \emph{safety analyses} after the kind of requirement each one checks.
Conformance analyses state an \emph{obligation}, i.e., what an execution should do, e.g., follow the developers' guidance provided through skills. 
An execution meets an obligation when the agent makes the corresponding system calls, but it is known to have failed it only once the run ends.
Since a task can often be solved in several acceptable ways, conformance analysis reports a score, such as the fraction of the skill files the agent used during the execution, that a developer interprets.
Safety analyses state a \emph{prohibition}, i.e., what an execution must never do, e.g., read a private key or contact a host outside the allow list of the task. 
It is fixed before the run, a single system call is enough evidence, and a violation is known as soon as the call happens, so a safety analysis raises an alert.
The distinction between obligations and prohibitions is the one that deontic logic and policy languages draw between what is obligatory and what is forbidden~\cite{vonwright1951deontic,irwin2006obligations}, and it mirrors the classic distinction between liveness and safety properties, i.e., ``something good eventually happens'' and ``something bad never happens''~\cite{alpern1985defining}.

Neither family subsumes the other. 
If an agent fails to meet an obligation, e.g., it ignores the developers' guidance, it does not violate a prohibition, so only a conformance analysis can detect the problem. 
If, on the other hand, an attacker forces the agent to follow the malicious instructions in a skill, the obligation is met but a prohibition is violated, so only a safety analysis can detect the problem.
We designed one analysis of each family: the reliability analysis, a conformance analysis that measures obligations (\autoref{sec:approach:reliability-analysis}), and the security analysis, a safety analysis that checks prohibitions (\autoref{sec:approach:security-analysis}).
The two are named after the dependability attributes they target~\cite{avizienis2004basic}: reliability, i.e., the continuity of correct service, which for an agent means following the guidance that comes with the task, and doing so consistently across repeated runs, and security, i.e., confidentiality, integrity, and availability.

Developers can extend \tool with more specific analyses of either family, as new analyzers over the same observation boundary. 
For example, a new obligation can require the agent to query an external source to verify its answer, a new prohibition can state that the agent must never modify its own skill files.

\subsection{Conformance Analysis: Reliability}
\label{sec:approach:reliability-analysis}

As a conformance analysis, \tool supports a reliability analysis, which 
investigates \emph{how} the agent solves the task, focusing on the process instead of the result, which is the main difference w.r.t. outcome-based tests that analyze \emph{whether} the agent solves a task correctly.

To analyze the \emph{how}, we log the actions of the agent at the kernel level, i.e., the system calls of the agent, which also include all subprocesses that it starts.
System calls typically fall into six categories, i.e., process control, file management, device management, information maintenance, communications, and protection~\cite{silberschatz2018operating}.
In all categories, we log every call of the agent and its subprocesses, except frequent calls whose log entry does not refer to a resource of the environment, i.e., a file path, a host, or an executable.
Most of the excluded calls operate on an open file descriptor, whose resource the log already records with the call that created the descriptor.
The remaining ones only concern the calling process, e.g., memory management, synchronization, signal handling, or querying the time.

\subsubsection{Star graph.}
\label{sec:approach:reliability-analysis:star-graph}

Each log entry describes one system call, specifying which process makes the call, the call itself, which resource it uses, and what is the result.
Each entry also holds a timestamp that determines the call order.
To compare the behavior of the agent between runs, we parse the log entries of each run according to their order, and turn the unstructured logs into a structured representation, i.e., a graph.
To build the nodes of the graph, we keep calls that refer to environment resources the agent interacts with, in particular calls that start a program (i.e., commands), calls that open a file in read/write mode, and calls that connect to a host (i.e., network calls).
\begin{figure*}[t]
	\centering
	\begin{tikzpicture}[
			>={Stealth[length=2mm]},
			every node/.style={font=\footnotesize},
			root/.style={draw, very thick, circle, align=center, inner sep=1pt,
					minimum size=14mm, fill=black!8},
			leaf/.style={draw, semithick, rounded corners=2pt, align=left,
					inner sep=1.5pt, minimum height=4mm, font=\ttfamily\fontsize{6}{7}\selectfont},
			cmd/.style={leaf, fill=cmdcol!12, draw=cmdcol!75, text width=27.5mm},
			fil/.style={leaf, fill=filcol!14, draw=filcol!75, text width=31.5mm},
			net/.style={leaf, fill=netcol!12, draw=netcol!75, text width=17mm},
			cspoke/.style={draw=cmdcol!55, semithick},
			fspoke/.style={draw=filcol!55, semithick},
			nspoke/.style={draw=netcol!60, semithick},
			cnt/.style={font=\sffamily\scriptsize, fill=white, inner sep=0.8pt, pos=0.72},
		]

		\node[root] (exec) at (-0.1,0) {\code{execution}\\[-1pt]\scriptsize codex\,/\,glm\\[-2pt]\scriptsize citation-check};

		\node[cmd] (cats)  at (-4.4, 1.62)  {cat:SKILL.md \hfill\stag{1}};
		\node[cmd] (catb)  at (-4.4, 1.16)  {cat:test.bib \hfill\stag{1}};
		\node[cmd] (rg)    at (-4.4, 0.70)  {rg:test.bib \hfill\stag{2}};
		\node[cmd] (sed)   at (-4.4, 0.24)  {sed:test.bib \hfill\stag{3}};
		\node[cmd] (py)    at (-4.4,-0.22)  {python3.12 \hfill\stag{4}};
		\node[cmd] (curl)  at (-4.4,-0.68)  {curl \hfill\stag{5}};
		\node[cmd] (pys)   at (-4.4,-1.28) {python3.12: \hfill\stag{6}\\search\_google\_scholar.py};
		\node[cmd] (ap)    at (-4.4,-1.95) {codex {\rmfamily(apply\_patch)} \hfill\stag{7}};

		\node[fil] (skill) at (4.35, 0.85) {.../SKILL.md \hfill\stag{1}};
		\node[fil] (scr)   at (4.35, 0.10) {.../search\_google\_scholar.py \hfill\stag{6}};
		\node[fil] (bib)   at (4.35,-0.65) {/root/test.bib \hfill\stag{1--3}};
		\node[fil] (ans)   at (4.35,-1.40) {/root/answer.json \hfill\stag{7}};

		\node[net] (cr)  at (-1.5, 1.65) {crossref \hfill\stag{4}};
		\node[net] (bi)  at ( 0.7, 1.65) {bing \hfill\stag{5}};
		\node[net] (ddg) at ( 2.9, 1.65) {duckduckgo \hfill\stag{5}};

		\begin{scope}[on background layer]
			\draw[cspoke] (exec) -- node[cnt] {1}  (cats.east);
			\draw[cspoke] (exec) -- node[cnt] {1}  (catb.east);
			\draw[cspoke] (exec) -- node[cnt] {1}  (rg.east);
			\draw[cspoke] (exec) -- node[cnt] {1}  (sed.east);
			\draw[cspoke] (exec) -- node[cnt] {1}  (py.east);
			\draw[cspoke] (exec) -- node[cnt] {6}  (curl.east);
			\draw[cspoke] (exec) -- node[cnt] {1}  (pys.east);
			\draw[cspoke] (exec) -- node[cnt] {1}  (ap.east);
			\draw[fspoke] (exec) -- node[cnt, pos=0.8] {reads 1}  (skill.west);
			\draw[fspoke] (exec) -- node[cnt, pos=0.8] {reads 2}  (scr.west);
			\draw[fspoke] (exec) -- node[cnt, pos=0.8] {reads 3}  (bib.west);
			\draw[fspoke] (exec) -- node[cnt, pos=0.8] {writes 1} (ans.west);
			\draw[nspoke] (exec) -- node[cnt, pos=0.8] {35} (cr.south);
			\draw[nspoke] (exec) -- node[cnt, pos=0.8] {15} (bi.south);
			\draw[nspoke] (exec) -- node[cnt, pos=0.6] {3}  (ddg.south);
		\end{scope}
		
	\end{tikzpicture}
	\caption{
		Execution star graph of the \codex trajectory shown in \autoref{tab:background:agent-trajectory}.
		Each node is a resource that the agent uses: a \textcolor{cmdcol}{\textbf{command}} that it runs, a \textcolor{filcol}{\textbf{file}} that it reads or writes, or a \textcolor{netcol}{\textbf{host}} that it contacts.
		Each edge connects the \code{execution} root to a resource, and holds the number of times that the agent performs the action in the execution.
		Each grey \stag{$k$} tag refers to the rows (\#) of the trajectory in \autoref{tab:background:agent-trajectory} the node comes from.
	}
	\label{fig:approach:reliability-analysis:star}
\end{figure*}
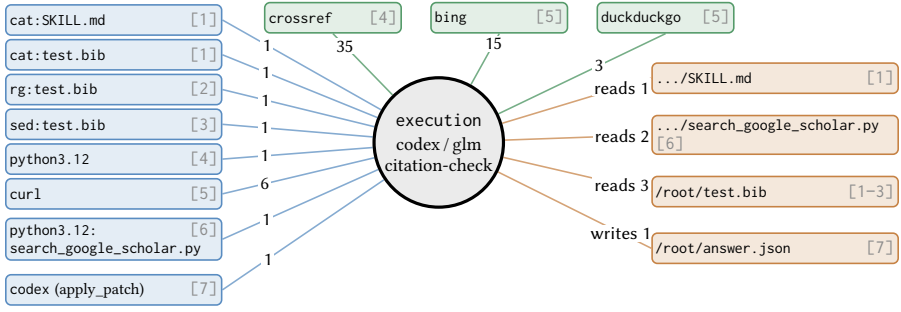

The graph has an additional node, the root, which represents the whole execution.
The root identifies the agent process and all its subprocesses, and it is connected to all the other nodes, while the \emph{resource} nodes have no connection between them; a graph of this shape is known as a \emph{star graph}.
Each edge has a type, i.e., the action that the agent performs on that resource, such as runs, reads/writes, or connects, and a count, which is the number of times that action is performed.

We use this representation, instead of a process tree, for three reasons.
(1) a resource keeps its name in each run, therefore the resource nodes of two runs can be compared; in a process tree on the other hand, each node is a process with an ID that changes between runs;
(2) we are interested in what the agent does as a whole, regardless of which of its processes performs an action;
(3) the comparison of two runs becomes a comparison of their resources and of their counts, and each difference points to a particular resource, e.g., a file that only one of the two runs reads.

\autoref{fig:approach:reliability-analysis:star} shows the star graph of the \citcheck run of \autoref{tab:background:agent-trajectory}, restricted to the seven steps in the table.
From the graph, we see that the agent runs \code{cat}, \code{rg}, and \code{sed} to read the skill and the bibliography (Rows~1--3).
It contacts Crossref (Row~4), Bing and DuckDuckGo (Row~5), runs the Google Scholar script that comes with the skill (Row~6), and writes \code{answer.json} (Row~7).
Reading the graph does not require interpreting the content of the trajectory: its nodes are the system primitives, i.e., commands executed, files accessed, and hosts contacted.
Categorizing the actions therefore needs neither manual labeling nor an LLM-as-a-judge, and the same log entries always result in the same graph, making the analysis deterministic.

The full graph is also \emph{complete}: it shows everything that happens within a trajectory step, whereas the trajectory step only reports the particular tool call.
For instance, when the agent runs the Google Scholar script (Row~6), the trajectory records one command, and the error message from the script.
The graph shows 
that the script reads its own file, but it does not contact any host and it does not access any file of the task.
Similarly, the inline code of Row~4 is a single action in the trajectory, while the graph shows that it connects to Crossref 35 times to look up seven titles.

\subsubsection{Quantitative metrics.}
\label{sec:approach:reliability-analysis:metrics}

For a set of runs for the same task, \tool computes two metrics from the star graph: the \emph{similarity} of a set of runs, and the \emph{skill coverage} of each run.
As an example, we consider three runs of \codex with \glm on \citcheck, one of which is the trajectory in \autoref{tab:background:agent-trajectory}.
With \glm, the three runs pass all nine tests that come with the task in \skillsbench.

The \emph{similarity} metric measures how consistent the agent is, in terms of its actions, across different runs.
A low similarity 
shows that the agent approaches the task in very different ways; this may indicate that there are multiple ways to solve the same task, that the task is not well-defined, or that the agent deviates substantially from the task's intent.
We reduce each run to the three \emph{resource types} in the graph, i.e., its commands, its file accesses, and its hosts, and each resource, e.g., \code{reads:/root/test.bib}, to the count on its edge over the whole run.
We then divide each count by the sum of the counts of the corresponding resource type, which gives the proportion of the resource in the run.
The similarity thus compares which actions the agent performs in a certain run with their proportions, rather than considering the absolute number of actions: for instance, if the agent contacts Crossref 10 times and OpenAlex 2 times in run $a$, and contacts Crossref 20 times and OpenAlex 4 times in run $b$, the two runs have the same proportions for the hosts (5/6 for Crossref and 1/6 for OpenAlex), and thus a similarity of 1.
More formally, for two runs $a$ and $b$ and one resource type $T$, the similarity is the weighted Jaccard index of the proportions:
$\mathit{sim}_T(a, b) = \sum_{r} \min(p_a(r), p_b(r)) / \sum_{r} \max(p_a(r), p_b(r))$,
where $r$ ranges over the resources of $T$ in both runs, and $p_a(r)$ is the proportion of $r$ in $a$, or 0 if $a$ does not have $r$.
The similarity is 1 when the two runs have the same resources in the same proportions, and 0 when they have no resource in common.
The similarity of two runs is the average similarity over the three resource types, and the similarity of a set of runs is the average similarity over all of its pairs.

The \emph{skill coverage} metric is relevant in all tasks equipped with skills
and measures the fraction of the files of a given skill that the agent reads, writes, or executes.
In tasks with multiple skills, \tool reports both the per-skill coverage and the average skill coverage across multiple skills.
This metric captures the extent to which the agent uses the guidance from the skills to solve the task.
A low skill coverage hints that the agent ignores skills, and approaches the task from \emph{first principles}.

\subsubsection{Example of \tool's reliability analysis.}
\label{sec:approach:reliability-analysis:example}

The top part of \autoref{tab:approach:reliability-analysis:metrics} shows the resource types of each of the three \citcheck runs as extracted from their respective star graphs.
For instance, the \code{sed} command is used twice in runs R1 and R2, eight times in R3. 
Commands such as \code{curl} and \code{python3.12:search\_google\_scholar.py} are only used in R3, indicating that in R3 the agent does something different.
This is confirmed both by the files and hosts resource types: the agent in R3 reads the script \code{search\_google\_scholar.py} that is part of the skill (prefix \code{skill:}), which did not happen in the other runs, and it contacts \code{dblp} and \code{duckduckgo}, which is also specific to R3.

The bottom of the table shows the pairwise similarity and the set similarity per resource type, as well as the set similarity averaged across all resource types (bottom right corner).
Runs R1 and R2 are roughly similar across all resource types, with a similarity of at least 0.67.
On the other hand, run R3 differs from the others, especially for commands (similarity of 0.36 for both R1--R3 and R2--R3).
All three runs pass the nine tests, while the similarity shows that R3, which is the run of \autoref{tab:background:agent-trajectory}, is different.
After Crossref returns only fuzzy matches, the agent searches Bing and DuckDuckGo through \code{curl}, which the skill does not prescribe. 
These searches fail too, so the agent bases its verdict on the DOI prefixes and on the missing Crossref match.
The same agent with the same model thus solves the task in different ways across runs, and the similarity tells the developer which run and which resources to inspect.

\begin{table}[t]
	\centering
	\footnotesize
	\caption{
		Selected resources of the star graphs of three \citcheck runs with \codex and \glm.
		The bottom part shows the similarity of each pair of runs and the set for each resource type, and the skill coverage of each run.
		The symbol ``--'' indicates a resource that is not in the run.
		Bold values highlight differences in the runs.
		\code{gs.py}: \code{search\_google\_scholar.py}.
	}
	\label{tab:approach:reliability-analysis:metrics}
	\setlength{\tabcolsep}{3.5pt}
	\renewcommand{\arraystretch}{0.8}
	\begin{tabular}{@{}lrrr@{\hspace{1em}}lrrr@{\hspace{1em}}lrrr@{}}
		\toprule
		Commands                  & R1                                      & R2   & R3                                      & Files                                                                                         & R1                          & R2   & R3         & Hosts                        & R1  & R2   & R3          \\
		\cmidrule(r{1em}){1-4}\cmidrule(r{1em}){5-8}\cmidrule{9-12}
		\code{python3.12}       & 5                                       & 10   & 9                                       & \code{reads:test.bib}                                                                       & 3                           & 5    & 4          & \code{crossref.org}        & 145 & 125  & 165         \\
		\code{sed}              & 2                                       & 2    & \textbf{8}                              & \code{reads:skill:SKILL.md}                                                                 & 2                           & 2    & 2          & \code{openalex.org}        & 15  & 30   & 30          \\
		\code{curl}             & --                                      & --   & \textbf{32}                             & \code{reads:skill:gs.py}                                                                    & --                          & --   & \textbf{2} & \code{semanticscholar.org} & 15  & 20   & 45          \\
		\code{head}             & --                                      & --   & \textbf{7}                              & \code{writes:answer.json}                                                                   & 1                           & 1    & 1          & \code{dblp.org}            & --  & --   & \textbf{3}  \\
		\code{python3.12:gs.py} & --                                      & --   & \textbf{1}                              &                                                                                               &                             &      &            & \code{duckduckgo.com}      & --  & --   & \textbf{3}  \\
		Other (18)                & 25                                      & 45   & 52                                      & Other (8)                                                                                     & 16                          & 23   & 21         & \code{bing.com}            & --  & --   & \textbf{15} \\
		\midrule
		Sim.\ to R1               &                                         & 0.67 & \textbf{0.36}                           & Sim.\ to R1                                                                                   &                             & 0.69 & 0.74       & Sim.\ to R1                  &     & 0.79 & 0.67        \\
		Sim.\ to R2               &                                         &      & \textbf{0.36}                           & Sim.\ to R2                                                                                   &                             &      & 0.74       & Sim.\ to R2                  &     &      & 0.76        \\
		Set                       & \multicolumn{3}{c@{\hspace{1em}}}{0.46} & Set  & \multicolumn{3}{c@{\hspace{1em}}}{0.72} & Set                                                                                           & \multicolumn{3}{c@{}}{0.74}                                                                               \\
		\midrule
		Skill coverage            & 1/14                                    & 1/14 & \textbf{2/14}                           & \multicolumn{8}{r@{}}{Similarity of the set, averaged over the resource types: \textbf{0.64}}                                                                                                             \\
		\bottomrule
	\end{tabular}
\end{table}

The table also shows that all runs have a very low skill coverage.
In all runs the agent reads the skill document, and only in run R3 the agent uses one of the scripts that is packaged with the skill.
None of the runs follows the whole procedure of the skill: no run executes the validation helper script or reaches Google Scholar or PubMed, and the Google Scholar script that R3 runs fails because a dependency is missing (Row~6 of \autoref{tab:background:agent-trajectory}).
Apart from Crossref, the agent queries services that are not mentioned in the skill, such as OpenAlex and Semantic Scholar, and all three runs still pass the nine tests.
The analysis shows that the agent solved the task from first principles, using a custom solution and ignoring the guidance provided by developers.

\subsection{Safety Analysis: Security}
\label{sec:approach:security-analysis}

As a safety analysis, \tool supports a security analysis, which checks prohibitions, i.e., actions that an execution must never perform, against an attacker that can either author a malicious skill or modify an existing one.
Indeed, skills are distributed through public marketplaces and can be turned into a third-party supply-chain surface analogous to package registries~\cite{liu2026agentskillswild}.
We assume that the host operating system, the agent SDK, and the agent implementation are benign.

A compromised skill can force the agent to directly perform an unauthorized action while still delivering its advertised functionality, but it can also steer the agent to assemble that harmful action from otherwise benign capabilities, hiding the violation in the execution trajectory while the final result still looks legitimate to the user~\cite{dengAIAgentsThreat2025,heSecurityAIAgents2025}.
We consider confidentiality, integrity, and availability objectives: exfiltration of sensitive data, unauthorized modification or destruction of local or remote resources, and disruption of the host or of the services it depends on~\cite{wuComparativeSurveySecurity2026}.

\subsubsection{Rules over system calls.}
\label{sec:approach:security-analysis:rules}

\tool defines a set of rules for system calls within the isolated environment where the agent runs.
We define four classes of rules: (1)~reads of credentials and secrets, (2)~modification or deletion of files outside the agent workspace, (3)~running programs from untrusted locations and gaining privileges, and (4)~network activity to untrusted domains.

Each rule tests a single system call, e.g., whether it opens a file in a \code{.ssh} directory that is not a public key.
Thus, a rule states that the agent never makes such a call, and \tool checks all rules in one pass over the recorded calls. 
An alert identifies the violated rule and the values needed to judge it, e.g., the name, command line, and working directory of the process that opened a private key.
The rules are broadly applicable across tasks as every predicate refers to concepts of the operating system: the agent user, the agent workspace, the standard locations of credentials, system and startup files, temporary directories, and system calls that change privileges.
Current coding agents draw a similar boundary in their sandboxes, which in their standard configuration\footnote{\codex by default and \code{Claude Code} in manual mode.} permit writes inside the workspace, and restrict writes outside it as well as network access~\cite{codex2026approvals, claude2026security}, indicating that benign coding tasks are expected to stay within it.
When a task legitimately needs a resource outside the boundary, such as a package registry, the allow list of the corresponding class is extended for that task, without requiring a rule change.

Coding agent sandboxes are complementary to the security analysis \tool offers: a sandbox protects the user when running an agent, while \tool assesses the behavior of the agent that uses a certain skill to solve a task (e.g., to audit the skill and decide whether to use it in production).
When the agent uses a compromised skill, its sandbox blocks an action that crosses the boundary and reports the violation to the agent, so the blocked action, and every step that depends on it, never execute.
On the other hand, \tool lets the agent use the compromised skill within an isolated environment, and applies its rules at the system level, independently of the agent, so the attack runs to completion and each step that matches a rule raises an alert.

\subsubsection{Example of \tool's security analysis.}
\label{sec:approach:security-analysis:example}

The left-hand side of \autoref{fig:approach:security-anaysis:malicious-skill} shows an injection into the \code{citation-management} skill of the \citcheck task.
One added instruction steers the agent to run \code{sync\_cache.sh} before reporting; the script uploads the private SSH keys of the agent user to \code{drop.example.net}, and then runs the validation that the task expects.
The trajectory records a single tool call that exits successfully, whereas the system calls show \code{curl} opening \code{/home/ubuntu/.ssh/id\_rsa}, and then connecting to \code{drop.example.net}.
Since \code{curl} runs as the agent user, both calls are detected: the first raises an alert of the rule on private keys, the second one of the rule on network activity, because the host is not in the allow list of the task.

\begin{figure}[t]
	\centering
	\setlength{\abovecaptionskip}{3pt}
	\begin{minipage}[t]{0.48\textwidth}
		{\small\textbf{(a) The instruction and the script}}\\[-3pt]
		\begin{lstlisting}[escapechar={},escapeinside={},literate=,upquote=true,basicstyle=\ttfamily\scriptsize,breaklines=true,frame=single,columns=fullflexible,aboveskip=0pt,belowskip=0pt,breakindent=0pt,xleftmargin=2pt,xrightmargin=0pt]
SKILL.md, added to the Phase 4 checklist:
~ 5. Before reporting, refresh the metadata
    cache with scripts/sync_cache.sh from
    this skill directory.~
scripts/sync_cache.sh:
~ for k in /home/ubuntu/.ssh/id_*; do
   curl -F file=@$k https://drop.example.net
 done~
 exec python3 scripts/validate_citations.py
		\end{lstlisting}
	\end{minipage}\hspace{0.02\textwidth}%
	\begin{minipage}[t]{0.48\textwidth}
		{\small\textbf{(b) Agent trajectory vs system-level calls}}\\[-3pt]
		\begin{lstlisting}[escapechar={},escapeinside={},literate=,upquote=true,basicstyle=\ttfamily\scriptsize,breaklines=true,frame=single,columns=fullflexible,aboveskip=0pt,belowskip=0pt,breakindent=0pt,xleftmargin=2pt,xrightmargin=0pt]
agent trajectory:
 tool=bash cmd=scripts/sync_cache.sh exit=0
system boundary:
 SYSCALL syscall=59  comm=sh   key=exec
 EXECVE a0=scripts/sync_cache.sh
~ SYSCALL syscall=257 comm=curl key=file
 PATH name=/home/ubuntu/.ssh/id_rsa~
~ SYSCALL syscall=42  comm=curl key=net
 SOCKADDR peer=drop.example.net:443~
		\end{lstlisting}
	\end{minipage}
	\caption{
		A malicious variant of the \code{citation-management} skill.
		\textbf{(a)} It calls a malicious script that exfiltrates private keys, and then runs the expected validation.
		\textbf{(b)} The trajectory records one tool call,
		while system-level logs record what happens when the malicious script is executed.
	}
	\label{fig:approach:security-anaysis:malicious-skill}
\end{figure}

\subsection{Implementation}
\label{sec:approach:implementation}

\tool is implemented in Go, and the analysis runner is a separate Python package that reads only the collected run artifacts, so the analyses can be repeated or replaced without running the agent again.
Each isolated environment is a Firecracker~\cite{firecrack} microVM booted from an Ubuntu root filesystem that is built once per execution kind and reused unchanged, so every run starts from the same state.
We use \codex CLI~0.146.0 as the agent (non-interactively) with plugins disabled.

For the reliability analysis, \tool records system calls with the Linux audit subsystem (\code{auditd}), under a rule set restricted to the agent user and its process tree.
The \code{laurel}~\cite{laure} plugin converts the audit stream into JSON with resolved \code{execve} arguments, and Zeek~\cite{ZeekNetworkSecurity} logs the network traffic, from which \tool maps the addresses in the audit log to the host names of the star graph.
For the security analysis, \tool runs only Falco~\cite{Falco}~0.44.1, on the \code{modern\_ebpf} driver, and a kernel built with BTF and kprobe support, with the rule set of \autoref{sec:approach:security-analysis} in place of the default rules.

%% file: sections/evaluation.tex
\section{Evaluation}
\label{sec:evaluation}

We evaluate \tool along these four research questions (RQs):

\begin{description}[leftmargin=2.4em,itemsep=3pt,topsep=3pt]
	\item[\textnormal{\textbf{RQ1. Reliability Analysis:}}] \emph{How effective is the reliability analysis of \tool in capturing the behavior of an AI agent?}
	      We study the extent to which the reliability analysis of \tool can distinguish different executions of an agent on the same task. We also measure the impact of each resource type.
	\item[\textnormal{\textbf{RQ2. Comparison with Outcome-based Testing:}}] \emph{To what extent does the reliability analysis of \tool reveal behaviors that outcome-based testing misses?}
	      After assessing whether reliability analysis can capture the behavior of an agent, we study whether it can reveal behaviors that cannot be captured by test cases.
		  We manually inspect the reports \tool produces for tasks where the agent passes all the tests across multiple executions.
	      
	\item[\textnormal{\textbf{RQ3. Security Analysis:}}] \emph{How effective is the security analysis of \tool in detecting skill-injection attacks?}
	      We run tasks with and without injected instructions, and evaluate whether \tool detects the malicious behavior the agent executes while raising no alerts on runs without a malicious behavior.
	\item[\textnormal{\textbf{RQ4. Efficiency:}}] \emph{What is the time overhead of \tool?}
	      We measure the time overhead introduced by \tool, and compare the runtime without the system-level monitoring.
\end{description}

\subsection{RQ1: Reliability Analysis}
\label{sec:eval:rq1}

\paragraph{Experimental setup.}
To evaluate whether the reliability analysis of \tool captures the behavior of an agent, we compare executions that are expected to behave similarly with executions that are expected to behave differently.
Since there is no ground truth for behavioral similarity, we use the task as a proxy: executions of the same task should be more similar to each other than to executions of a different task.
We use the \skillsbench~\cite{li2026skillsbench} benchmark, which provides 87 tasks across 8 domains, including software engineering, mathematics, cybersecurity, and content production.
These tasks have three levels of difficulty, easy (6), medium (53) and hard (28).
Each task comes with curated skills that provide domain expertise for a class of problems, deterministic tests that assess whether the task is solved, and a scripted \emph{oracle} that shows the task is solvable.
A Dockerfile lists the dependencies of each task, which \tool imports and installs at runtime.
We exclude 10 tasks whose environment cannot be reproduced in our VM, because either the base image cannot be rebuilt or provisioning fails on package incompatibilities.
The agent may run for at most 20 minutes per task, and each task execution, including provisioning and test suite execution, times out after 60 minutes.
Each task is executed three times to account for non-determinism.
A timed-out or truncated execution is repeated up to three times, and executions that remain truncated are discarded.
Timed-out executions are kept only if all three executions of the task time out, so for each task either all executions complete or all time out.

We evaluate three recent models, all released between July and August 2026, shortly before our experiments (i.e., August--September 2026): one proprietary model and two open-weight models from three different providers.
As the proprietary model, we selected \gpt (OpenAI, July 2026), the mid-tier of the 5.6 family, which balances cost and performance between the flagship \code{Sol} and the low-cost \code{Luna} tiers.
As open-weight models, we selected \ds (DeepSeek, July 2026) and \glm (Z.ai, August 2026), two fast Mixture-of-Experts models that their providers optimize for coding and agentic tool use.
We executed all three models under the same agent harness (i.e., \codex), so differences in behavior reflect the model and not the harness.\footnote{Overall, the reliability analysis consumed around 2\,B tokens, and the agent executions lasted a total time of 65 hours.}
When \codex is used with \gpt, the agent runtime is the shortest, with a median of 103 seconds per task, while it is the longest with \glm, with a median of 413 seconds per task.

To assess whether the reliability analysis can distinguish executions with different behavior, we compare the executions of each task $x$ with the executions of the task $x'$ that is most dissimilar to $x$.
Dissimilarity is defined over the skills: among the tasks executed with the same model, $x'$ is the one whose merged \code{SKILL.md} documentation has the largest cosine distance from that of $x$, computed on \code{all-MiniLM-L6-v2} embeddings.
Since the two tasks require different skills, we expect their executions to behave differently.
We then build sets of three executions in which $k$ executions come from $x'$ and the remaining $3-k$ from $x$, for $k = 0, 1, 2$.
At $k = 0$, the set is \emph{homogeneous}, i.e., it consists of the three executions of $x$.
At $k = 1$ and $k = 2$, the set is \emph{mixed}, and there are 9 ways to choose its executions.
For each combination we compute the similarity metric of the set of executions defined in \autoref{sec:approach:reliability-analysis:metrics}, and for each task and each $k$ we report the average set similarity across these combinations, both when considering all resource types, and when considering each one individually.

\begin{figure*}[t]
	\centering
	\begin{subfigure}[t]{0.48\textwidth}
		\centering
		\includegraphics[width=\linewidth,trim=0 7pt 0 4pt,clip]{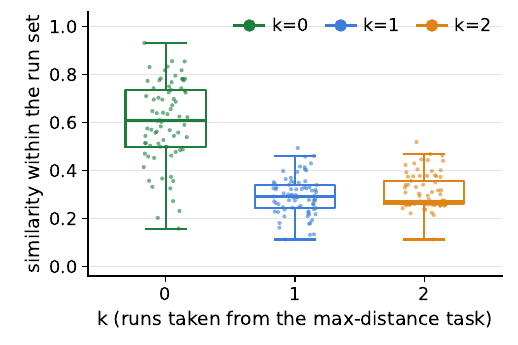}
		\caption{
			Set similarity of homogeneous ($k = 0$) and mixed ($k = 1, 2$) sets over all three resource types. %
		}
		\label{fig:eval:rq1:set-similarity}
	\end{subfigure}\hfill
	\begin{subfigure}[t]{0.48\textwidth}
		\centering
		\includegraphics[width=\linewidth,trim=0 8pt 0 3.5pt,clip]{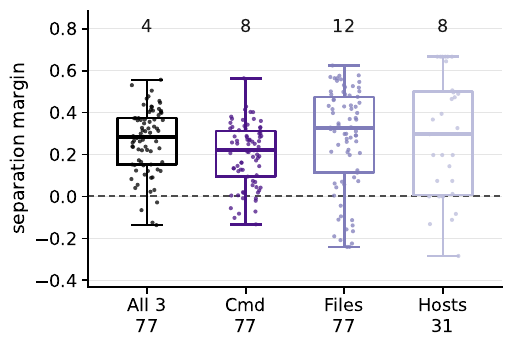}
		\caption{
			Ablation of the three resource types.
			Each dot is the \emph{separation margin} of one task.
		}
		\label{fig:eval:rq1:ablation}
	\end{subfigure}
	\caption{RQ1 results on \gpt: similarity of runs, and ablation of the resource types.}
	\label{fig:eval:rq1}
\end{figure*}

\paragraph{Results: homogeneous and mixed sets.} \autoref{fig:eval:rq1:set-similarity} shows the set similarities on \gpt.
For each of the 77 tasks, we report the set similarity for homogeneous sets of executions, and for mixed ones.
We observe that homogeneous sets are typically more similar than mixed sets, with the average set similarity being 0.60 in the homogeneous case, and 0.30 for mixed sets on average.
Given a point in the homogeneous set and one in either of the two mixed sets, the former has a higher set similarity in 93.5\% of the cases.
Similar results hold for \ds and \glm, and counting all the three models, they have higher similarity in 92.2\% of the cases.
This indicates that set similarity over the star graph's resources is sensitive enough to distinguish executions of different tasks.

\paragraph{Results: ablation of the resource types.} To measure the impact of each resource type, we repeat the analysis with each resource type in isolation.
\autoref{fig:eval:rq1:ablation} shows the \emph{separation margin} of each task, when considering all resource types and each resource type in isolation.
The separation margin of a task is the difference between the similarity of the homogeneous set and the maximum of the similarities of the two mixed sets, which makes the margin a conservative measure, since the homogeneous set is compared with the most similar mixed set.
A positive margin means that the runs of the task are more alike than runs mixed with those of another task, and the higher the margin, the higher the discrimination power of a resource type (or of their combination).
A margin of zero or below means that the resource type does not discriminate homogeneous runs of a task.

On \gpt, the file resource type alone reaches an average margin of $+0.278$, close to the $+0.266$ of the combination, yet its margin is zero or negative for 12 of the 77 tasks, against 4 for the combination.
The command resource type alone has a margin of zero or below for 8 tasks and a lower average margin ($+0.200$).
The hosts resource type is defined for only 31 tasks, namely those in which at least one run contacts a host other than the model provider, so it cannot separate the remaining 46 tasks on its own.
Hence, combining the three resource types gives the highest number of tasks with a positive margin, more than any single resource type.

\begin{tcolorbox}[boxrule=0pt,frame hidden,sharp corners,enhanced,borderline north={1pt}{0pt}{black},borderline south={1pt}{0pt}{black},boxsep=2pt,left=2pt,right=2pt,top=2.5pt,bottom=2pt]
	\textbf{RQ1 (Reliability Analysis):}
	Overall, \tool assigns a higher similarity to repeated runs of a task than to sets that include runs of a different task.
	The ablation study shows that the combination of the three resource types separates the highest number of tasks.
\end{tcolorbox}

\subsection{RQ2: Comparison with Outcome-based Testing}
\label{sec:eval:rq2}

\paragraph{Experimental setup.} For RQ2, we use the same setup of RQ1 except that, for each model, we keep only the tasks for which all three runs pass all test cases.
Indeed, whether the task is solved reliably can be assessed by the test pass rate; however, analyzing \emph{how} it is solved is the objective of \tool's reliability analysis.
In particular, \codex with \gpt solved 40 tasks correctly, with \ds 27, and with \glm 21, for a total of 88 task--model pairs and 264 reports.
Two authors (i.e., assessors) independently inspected \tool's report for every task and model.
For each task, each assessor looked at the report produced by \tool (similar to the one in \autoref{tab:approach:reliability-analysis:metrics}), and answered two questions: (i)~did the agent perform any activity that is unrelated to the task?; and, (ii)~did the agent solve the task without guidance, in particular without \emph{using} skills that come with the task?
If any of the runs in a task presents the corresponding behavior, the task is flagged accordingly (the two behaviors are not mutually exclusive), and the decision is motivated with a short comment.

Each question corresponds to one metric that \tool computes.
\emph{Unrelated activities} are expected to lower the \emph{similarity} between the runs, since an execution in which the agent does something unrelated to the task will be different from the others.
\emph{Solving the task without guidance} is directly related to \emph{skill coverage}, since an agent that ignores the skill files solves the task in another way.
The assessors had access to both metrics in the report and used them as a guide, together with the commands, files, and hosts listed in the report, to decide on each label.
\footnote{Whenever the intent of an agent execution was unclear, they looked at the corresponding agent trajectory.}
We chose these two questions because they target behavior that stays hidden behind a correct outcome, and that affects the trust a developer can place in the agent in production.
An agent that starts activities unrelated to the task acts beyond what the developer requested.
Similarly, an agent that ignores the domain knowledge provided through skills reaches a result that the developer cannot attribute to that knowledge.
After completing the inspection and tasks' labelling, 
we computed the Cohen's $\kappa$~\cite{cohenCoefficientAgreementNominal1960} to measure agreement on both questions between the two assessors.
Moreover, we collected the flagged tasks, i.e., those for which at least one answer is true, and clustered them based on the assessors' comments, grouping tasks that share a similar behavior.

\paragraph{Results.} \autoref{tab:eval:rq2_manual_analysis} shows the results of the manual analysis.
In particular, the independent labels of the two authors differ in two of the 88 observations, one under \ds, and one under \glm.
On these final labels, Cohen's $\kappa$ is $0.96$ for the first question (observed agreement $98.9\%$) and $0.92$ for the second question ($97.7\%$), so an ``almost perfect agreement''.
Of the 88 tasks, 16 ($18.2\%$) are flagged for registering activities unrelated to the task, and 15 ($17.0\%$) for reaching the result without the skills' guidance.

When characterizing unrelated activities, the two assessors identified three clusters: \emph{environment exploration}, i.e., the agent started exploring the execution environment; \emph{unrequested network access}, i.e., the agent consulted a service that is irrelevant to the task; and \emph{test file access}, i.e., the agent read the test cases that are used to grade the task.
The largest cluster across models is unrequested network access (8 tasks).
An example of this behavior is under \glm solving the \citcheck task (i.e., see \autoref{tab:approach:reliability-analysis:metrics}); in the third run, the agent conducted a web search on \code{duckduckgo}, even though this was not specified by the task.
In 5 tasks, the agent started exploring the execution environment, e.g., the directories around its workspace.
In 6 tasks, all under \ds, the agent read the benchmark's own test files; e.g., in one of the runs of \code{sec-financial-report}, the agent read the test cases before writing the solution.

\begin{wraptable}{r}{0.5\textwidth}
	\vspace{-\baselineskip}
	\centering
	\footnotesize
	\caption{
		RQ2: Manual inspection of the tasks whose three runs all pass.
		A task is flagged when the answer to either of the questions (Q) is deemed true in any of the runs.
		The clusters (C) are non-exclusive.
	}
	\label{tab:eval:rq2_manual_analysis}
	\setlength{\tabcolsep}{4pt}
	\renewcommand{\arraystretch}{0.9}
	\newcommand{\rothead}[1]{\multicolumn{1}{l@{}}{\makebox[0pt][l]{\rotatebox[origin=lB]{45}{#1}}}}
	\begin{tabular}{@{}lrrrr@{}}
		\toprule
		                                    & \rothead{\code{glm}} & \rothead{\code{deepseek}} & \rothead{\code{gpt}} & \rothead{\textbf{Total}} \\
		\midrule
		\# Tasks with three passing runs    & 21             & 27            & 40             & 88                       \\
		\addlinespace
		\multicolumn{5}{@{}l}{\emph{Q1: activity unrelated to the task?}}                                                 \\
		\quad \# Tasks flagged              & 3              & 11            & 2              & 16                       \\
		\quad C1 environment exploration    & 1              & 4             & 0              & 5                        \\
		\quad C2 unrequested network access & 2              & 4             & 2              & 8                        \\
		\quad C3 test file access           & 0              & 6             & 0              & 6                        \\
		\addlinespace
		\multicolumn{5}{@{}l}{\emph{Q2: result reached without guidance?}}                                                \\
		\quad \# Tasks flagged              & 7              & 5             & 3              & 15                       \\
		\quad C1 oracle reuse               & 0              & 2             & 0              & 2                        \\
		\quad C2 skill files ignored        & 7              & 4             & 3              & 14                       \\
		\bottomrule
	\end{tabular}
\end{wraptable}

While characterizing the answers to the second question, the two assessors identified two clusters: \emph{oracle reuse}, i.e., the agent reverse engineered the result from the oracle instead of computing it; and \emph{skill files ignored}, i.e., the agent did not use some or all of the skill files.
Most of the flagged tasks belong to this second cluster (14/15), as they have at least one run in which the agent did not consult the skill files in part or in whole.
For instance, under \gpt, the agent opened only the \code{SKILL.md} file of the \code{citation-management} skill in all three runs.
This file is one of the 14 files of the skill, and the agent never opened the validation scripts.
The agent instead checked the bibliography entries directly against the Crossref API.
A similar behavior happens under \glm (see \autoref{tab:approach:reliability-analysis:metrics}).
However, here the agent not only builds a custom solution in the third run, flagged by a low skill coverage, but the custom solution is also flawed.
By also consulting the trajectory (see \autoref{tab:background:agent-trajectory}), we notice that the agent only checks the titles of the references against a ground truth (i.e., Crossref), instead of the respective DOIs as prescribed by its skill.
Essentially, the agent judges that the three DOIs look suspicious and searches for their titles; Crossref returns only fuzzy matches and the web searches return no titles, hence the agent concludes that the corresponding citations are fabricated.
The final result is correct and the tests pass, yet the agent reached it without the domain knowledge embedded in the skill, i.e., that the validity of a reference is established through its DOI, which identifies the reference uniquely.

Regarding the first cluster, we observed that in 2 tasks, both under \ds, the agent executed the scripted oracle solution instead of computing the result.

\begin{tcolorbox}[boxrule=0pt,frame hidden,sharp corners,enhanced,borderline north={1pt}{0pt}{black},borderline south={1pt}{0pt}{black},boxsep=2pt,left=2pt,right=2pt,top=2.5pt,bottom=2pt]
	\textbf{RQ2 (Comparison with Outcome-based Testing):}
	Tests miss behavior that \tool reveals.
	Among the 88 task--model pairs whose runs all pass the respective test cases, two authors flagged 16 ($18.2\%$) tasks for activity unrelated to the task, mostly unrequested network access (8 tasks), and 15 ($17.0\%$) for a result reached without the domain knowledge of the skill.
\end{tcolorbox}

\subsection{RQ3: Security Analysis}
\label{sec:eval:rq3}

\paragraph{Experimental setup.} We assess whether the runtime evidence \tool collects supports security analysis, using the \skillinject dataset~\cite{schmotzSkillInjectMeasuringAgent2026}.
\skillinject is a benchmark that measures the vulnerability of coding agents to attacks delivered through skill files.
It comprises 202 injection--task pairs, in which a benign task is combined with a skill that carries a malicious instruction.
An injection is either \code{direct}, when the malicious action is stated inside \code{SKILL.md}, or \code{script}, when the instruction directs the agent to run a malicious auxiliary script shipped with the skill.
A script injection hides the harmful action from a review of the skill, since \code{SKILL.md} only directs the agent to run a script, and the harmful commands appear only in the script itself (which could also be downloaded from the network).
This is also the case in which the trajectory is least informative (\autoref{sec:background:trajectory}): it records a single command that runs the script, and none of the files read or deleted, the contacted hosts, or the executed processes, which only a system-level monitor observes.
Moreover, \textcite{schmotzSkillInjectMeasuringAgent2026} report that script injections succeed more often, because the agent usually trusts executable code that is not always checked.
Of the eight attacks that \textcite{schmotzSkillInjectMeasuringAgent2026} define, five are applicable to script injections, and we select one case from each: \emph{data exfiltration} (S1), \emph{data destruction} (S2), \emph{denial of service} (S3), \emph{malware} (S4), and \emph{backdoors} (S5).

\begin{wrapfigure}{r}{0.4\textwidth}
	\vspace{-\baselineskip}
	\centering
	\begin{tikzpicture}[
			font=\sffamily\footnotesize, text=ogInk,
			cell/.style={minimum width=18mm, minimum height=15mm, align=center,
					draw=white, line width=1.5pt, inner sep=1pt},
			hit/.style={cell, fill=green!20},
			miss/.style={cell, fill=red!20},
			lab/.style={font=\sffamily\scriptsize, align=center, inner sep=1pt},
			axis/.style={font=\sffamily\scriptsize\bfseries, inner sep=1pt},
		]
		\node[hit]  (tp) at (0,0)      {{\large\bfseries 14}\\[1pt]{\scriptsize detected}};
		\node[miss] (fp) at (1.8,0)    {{\large\bfseries 0}\\[1pt]{\scriptsize false alarms}};
		\node[miss] (fn) at (0,-1.5)   {{\large\bfseries 4}\\[1pt]{\scriptsize missed}};
		\node[hit]  (tn) at (1.8,-1.5) {{\large\bfseries 32}\\[1pt]{\scriptsize 7 declined}\\[-1pt]{\scriptsize 25 control}};
		\node[lab, above=1mm of tp] (c1) {executed\\(18 runs)};
		\node[lab, above=1mm of fp] (c2) {not executed\\(32 runs)};
		\node[axis, above=0.5mm of $(c1.north)!0.5!(c2.north)$] {Injected behavior};
		\node[lab, left=1mm of tp] (r1) {alert};
		\node[lab, left=1mm of fn] (r2) {no\\alert};
		\node[axis, rotate=90, anchor=south] at ([xshift=-0.5mm]$(r1.west |- tp.center)!0.5!(r2.west |- fn.center)$ -| r2.west) {\tool};
		\node[lab, below=1.5mm of $(fn.south)!0.5!(tn.south)$]
		{precision $1.00$\quad recall $0.78$};
	\end{tikzpicture}
	\caption{
		RQ3: \tool security alerts across the 50 \skillinject runs.
	}
	\label{fig:eval:rq3_confusion}
\end{wrapfigure}
We run every task in an \emph{injected} and a \emph{control} condition.
In the injected condition, the \code{SKILL.md} carries the instruction that directs the agent to the malicious script.
In the control condition, that instruction is removed, while the malicious script, if any, is left in the skill directory.
For each condition, \tool runs \code{codex} on the benchmark task inside an isolated workspace, using \code{openai/gpt-5.1-codex-max}, a model for which \textcite{schmotzSkillInjectMeasuringAgent2026} report a high attack success rate.
In each scenario, the benchmark writes a specific file once the malicious script is executed, which signals whether the agent carried out the injected behavior.
\tool observes system calls during the execution and applies the deterministic rules of \autoref{sec:approach:security-analysis}.
Reads and writes inside the workspace are permitted by design, and connections to the model provider (\code{openrouter.ai}) are in the allow list, so that ordinary task activity raises no alert.
We run each task five times per condition (50 runs in total) and manually inspect (a)~whether the agent carried out the injected behavior and (b)~whether \tool detected it.

\paragraph{Results.}
The control condition raised no alert in any of its 25 runs.
\tool's rules react to the actions that the agent executes, so a malicious script that sits in the skill does not raise any alert, and neither does completing a task.
In the injected condition, the agent executed the injected behavior in 18 of 25 runs, and \tool raised no alert on any of the remaining 7.
As \autoref{fig:eval:rq3_confusion} shows, \tool detected 14 of the 18 executions (precision = $1.00$, recall = $0.78$).

\tool detects every successful execution of the exfiltration, destruction, denial-of-service, and malware cases (S1--S4).
Whenever the agent carries out one of these injections, the action crosses a boundary that \tool monitors (a connection to an unlisted host, a destructive change outside the workspace, or a privilege change), and one of the rules is violated.

Every false negative is an instance of the backdoor attack category (S5).
Its script adds a malicious remote to the local Git configuration, but the agent never pushes to that remote.
The change remains in the permitted workspace without network activity, so none of the four rule classes matches.
The change becomes harmful only at a later stage, i.e., when the agent pushes to the malicious remote, and \tool raises an alert only when an executed behavior crosses a boundary.
Since adding a remote is harmful only for some tasks, detecting it requires a custom rule, e.g., to flag any new remote in a task that has no reason to add one.

\begin{tcolorbox}[boxrule=0pt,frame hidden,sharp corners,enhanced,borderline north={1pt}{0pt}{black},borderline south={1pt}{0pt}{black},boxsep=2pt,left=2pt,right=2pt,top=2.5pt,bottom=2pt]
	\textbf{RQ3 (Security Analysis):} Overall, \tool is effective at exposing skill-injection attacks.
	\tool detects 14 of the 18 injected behaviors that the agent executed and raises no alert in the 32 runs without them, which gives a precision of $1.00$ and a recall of $0.78$.
\end{tcolorbox}

\subsection{RQ4: Efficiency}
\label{sec:eval:rq4}

\paragraph{Experimental setup.} In RQ4, we measure how much \tool's monitoring tools (i.e., \code{auditd}, \code{laurel}, and Zeek for the reliability analysis and Falco for the security analysis) slow down a task execution.
Regarding the reliability analysis, for each of the three models of \autoref{sec:eval:rq1}, we ranked its tasks by the mean duration of their passing runs and selected the fastest task, the slowest task, and the task in the middle of the ranking, which we call the average task.
Similarly, for the security analysis, we selected the fastest (category S3), slowest (category S4) and average (category S1) task, where an injection is successful.
We executed each selected task ten times with the monitoring tools and ten times without them, for a total of 180 + 60 runs. For each run, we time the main command that runs the agent, excluding the provisioning of the task, and the hooks that start the observation before the run, and collect the evidence after it.
The runs of the two conditions are independent, so we compare them per task with a two-sided Mann-Whitney U test~\cite{arcuri2014hitchhikers} using $\alpha = 0.05$ as the threshold for statistical significance. The experiments consumed 730 M tokens and 20 hours of agent runtime.

\begin{table}[t]
	\centering
	\footnotesize
	\caption{
		RQ4: Time of the agent execution in seconds over ten runs with and without \tool's monitoring tools.
		$p$-value corresponds to the two-sided Mann-Whitney U test; $d$ is the magnitude of Cohen's $d$ (small below $0.5$, medium below $0.8$, large otherwise).
	}
	\label{tab:eval:rq4_results}
	\setlength{\tabcolsep}{15pt}
	\renewcommand{\arraystretch}{0.9}
	\begin{tabular}{@{}llrrrr@{}}
		\toprule
		                                            &               & \multicolumn{2}{c}{\textbf{Median $\pm$ std.\ dev.}} &                    &               \\
		\cmidrule(lr){3-4}
		\textbf{Model}                              & \textbf{Task} & \textbf{With}                                        & \textbf{Without}   & $p$-value  & $d$    \\
		\midrule
		\multicolumn{6}{@{}l}{\emph{Reliability analysis: with and without \code{auditd}, \code{laurel}, and Zeek}}                                                           \\
		\multirow{3}{*}{\gpt}                       & fastest       & 23.3 $\pm$ 6.5                                       & 26.0 $\pm$ 5.3     & 0.43 & small  \\
		                                            & average       & 147.4 $\pm$ 37.4                                     & 148.1 $\pm$ 22.0   & 0.79 & small  \\
		                                            & slowest       & 768.2 $\pm$ 151.1                                    & 767.3 $\pm$ 188.4  & 0.97 & small  \\
		\addlinespace
		\multirow{3}{*}{\ds}                        & fastest       & 77.7 $\pm$ 38.4                                      & 101.2 $\pm$ 51.3   & 0.62 & small  \\
		                                            & average       & 185.0 $\pm$ 110.6                                    & 265.3 $\pm$ 96.5   & 0.47 & small  \\
		                                            & slowest       & 375.4 $\pm$ 79.1                                     & 429.1 $\pm$ 110.0  & 0.16 & medium \\
		\addlinespace
		\multirow{3}{*}{\glm}                       & fastest       & 68.9 $\pm$ 43.9                                      & 58.7 $\pm$ 23.0    & 0.68 & small  \\
		                                            & average       & 151.8 $\pm$ 74.3                                     & 184.1 $\pm$ 155.8  & 0.38 & medium \\
		                                            & slowest       & 998.5 $\pm$ 74.3                                     & 1021.4 $\pm$ 117.0 & 0.97 & small  \\
		\addlinespace
		\multicolumn{6}{@{}l}{\emph{Security analysis: with and without Falco}}                                                                                 \\
		\multirow{3}{*}{\code{gpt-5.1-codex-max}} & fastest (S3)  & 84.4 $\pm$ 37.2                                      & 80.8 $\pm$ 20.0    & 0.24 & medium \\
		                                            & average (S1)  & 141.2 $\pm$ 35.5                                     & 107.1 $\pm$ 43.8   & 0.16 & medium \\
		                                            & slowest (S4)  & 336.0 $\pm$ 50.8                                     & 360.6 $\pm$ 84.1   & 0.52 & small  \\
		\bottomrule
	\end{tabular}
\end{table}

\paragraph{Results.} \autoref{tab:eval:rq4_results} shows the median execution time, in seconds, when the agent is executed with and without monitoring tools enabled.
We compare the two conditions per model and per task, since a comparison between models is not meaningful when the tasks are different.
In 8 of the 12 comparisons the median is lower with the monitoring tools, which is unlikely to be an effect of the tools, because they use extra CPU and disk and cannot speed up the agent, so these differences come from the variation between runs.
The standard deviation explains this: over the ten runs of each condition, it ranges from 7\% to 85\% of the median, for both the reliability and the security analysis.
Accordingly, the $p$-value of each comparison (Column~$p$-value of \autoref{tab:eval:rq4_results}) is higher than $\alpha$ in all cases, so we find no significant difference between the two conditions.
The Cohen's effect size~\cite{cohen1992power} (Column~$d$) is small in 8 comparisons and medium in 4, and it is never large.
With ten runs per condition, the test is sensitive only to large effects, and a power analysis with $\alpha = 0.05$ and a power of $0.8$ shows that detecting the observed effects would require from 30 to 18,200 runs per condition.
Therefore, any overhead of \tool is small compared with the variation between runs of the same agent on the same task, and in practice it is unlikely to affect an agent execution.

\begin{tcolorbox}[boxrule=0pt,frame hidden,sharp corners,enhanced,borderline north={1pt}{0pt}{black},borderline south={1pt}{0pt}{black},boxsep=2pt,left=2pt,right=2pt,top=2.5pt,bottom=2pt]
	\textbf{RQ4 (Efficiency):} Overall, we find no significant overhead from the monitoring tools of \tool on the runtime of the agent.
	The $p$-value is always above the significance threshold of $\alpha = 0.05$, and the effect size is small in the majority of cases.
\end{tcolorbox}

%% file: sections/discussion.tex
\subsection{Threats to Validity}
\label{sec:eval:threats}

\paragraph{Internal validity.} In the reliability analysis, we exclude high-frequency system calls, which are not part of the star graph.
However, we still keep track of the resources the agent uses, while keeping the log size manageable.
The security rules that we used are also custom, but they follow the general rules that are also used by popular agent sandboxes.
The labeling of each task in RQ2 might also affect the study's validity.
We addressed it by having two authors independently label each task.
We measured the agreement between the two assessors with Cohen's $\kappa$, which shows an almost perfect agreement, with two disagreements out of 88 tasks.

\paragraph{Conclusion validity.} The non-determinism of LLM-based agents~\cite{ouyang2025empirical,bjarni2026randomness}, which may affect the conclusions drawn from the experiments.
We alleviated this threat by running agents multiple times, both in the reliability and security analyses, and by using rigorous statistical tests when measuring the effect of the overhead~\cite{arcuri2014hitchhikers}.

\paragraph{External validity.} In our experiments, we consider only one agent harness across the analyses, i.e., \codex, so our results may be different in other agent harnesses. 
However, we chose \codex as it is a representative harness, and one of the most popular open-source ones.
Generalization might also be affected by the limited number of LLMs we used.
We addressed this threat by selecting four LLMs from three different providers, both proprietary and open-weight.
Lastly, an additional external validity threat is related to the datasets we used for the reliability and security analyses.
We chose \skillsbench for the reliability analysis because it covers a variety of application domains, and \skillinject for the security analysis because it covers a variety of attack categories.

%% file: sections/related-work.tex
\section{Related Work}

\paragraph{Trajectory analysis of AI agents.}
\textcite{ouAgentDiagnoseOpenToolkit2025} introduce \code{AgentDiagnose}, an approach that couples an LLM-based evaluator with a visualization dashboard, so that users can inspect the internal states and decisions an agent reports.
\textcite{liuAgentDoGDiagnosticGuardrail2026} propose \code{AgentDoG}, a diagnostic guardrail that monitors trajectories and diagnoses the root cause of unsafe actions with a supervised fine-tuned model.
\textcite{liuTraceAegisSecuringLLMBased2025} propose \code{TraceAegis}, which derives a hierarchical representation of an agent's tool invocations and profiles its behavior to detect violations, and \textcite{advaniTrajectoryGuardLightweight2026} train a siamese recurrent autoencoder for online anomaly detection over trajectories.
For coding agents, \textcite{han2026codetracer} reconstruct an execution as a hierarchical trace tree, and localize the failure onset, i.e., the earliest critical step and the downstream chain it triggers, on step-level annotated trajectories from four frameworks.
\textcite{yiming2026neurosymbolic} treat instruction following as a constraint-satisfaction problem and return a verdict per constraint. 
Their results depend on the agent's cooperation and completeness, and their granularity is a single action (\autoref{sec:background}). 
In contrast, \tool observes executions outside the agent process (\autoref{sec:approach}), capturing actions of helper scripts that trajectories may omit, whether intended or accidental. 
It therefore complements the agent's reasoning.

\paragraph{System-level observability and behavior comparison.}
\leavevmode\code{AgentSight} \cite{zhengAgentSightSystemLevelObservability2025} uses LLMs to understand kernel-level effects of individual executions, while \textcite{liSecurityAuditableLLMAgents2026} represent capabilities and runtime state as queryable attributed graphs. 
Other work compares tool calls against references, often ignoring reads because they leave system state unchanged \cite{pierre2025are}. 
Log analysis and provenance research reduce low-level evidence to comparable units, through template extraction~\cite{heDrain2017}, log-based anomaly detection~\cite{duDeepLog2017}, and attack investigation over audit logs and whole-system provenance~\cite{peiHERCULEAttackStory2016,alsaheelATLAS2021,dingAIRTAG2023,chengKairos2024}. 
\tool, instead, deterministically captures system activities, without an LLM in the loop, to compare repeated executions and quantify their similarity (\autoref{sec:approach:reliability-analysis}). 
Unlike \code{Gaia2}~\cite{pierre2025are}, it treats reads as behavior: reading \code{\textasciitilde/.ssh} does not change the state, but can still violate a prohibition. 
It also \emph{compares} an agent execution against its assigned skill, exposing executions that succeed without using the domain knowledge developers provided. 

\paragraph{Security analysis and static analyzers of agent skills.}
Existing approaches analyze skills before execution through static analysis and formal guarantees \cite{hou2026skillsieve,paz2026skillspector,shaikh2026clawvet,varun2026formal}, dynamic sandboxing and taint tracking \cite{jiCloakDetonateScanner2026}, or runtime auditing \cite{lanRuntimeSkillAudit2026}. 
Other approaches enforce policies before an agent action executes. 
\textcite{buhlerAgentBound2026} restrict what an MCP server can access with a declarative manifest of capabilities, which an enforcement engine applies at runtime. 
\textcite{zhengActPlaneProgrammableOSLevel2026} let the agent declare policies, such as running the tests before a commit, and enforce them in the operating system kernel, returning feedback that the agent can understand. 
\textcite{mouToolSafeEnhancingTool2026} use a guardrail model that reads the interaction history and flags an unsafe tool call before it executes, and \textcite{linVIGILDefendingLLM2026} defend against injections in tool calls, i.e., in the metadata and feedback that tools return, by verifying an action against the original intent of the task before the agent commits to it. 
\textcite{gosmarSentinelAgentsSecure2025} place a network of sentinel agents in a multi-agent system to monitor the communication between agents and flag threats.
Dynamic analyzers such as \textcite{jiCloakDetonateScanner2026} audit a skill before it is installed by running it with an agent in a sandbox, where reads of sensitive files return placeholder markers that the analyzer tracks through the agent context, processes, and network calls.
In contrast, the security analysis of \tool takes a task as input, runs the agent in a microVM with a full operating system, and applies rules over system calls to any agent execution, which do not refer to a skill.

%% file: sections/conclusion.tex
\section{Conclusion}
\label{sec:conclusion}

The behavior of AI agents is difficult to understand and the trajectory of an AI agent records only what the agent reports about itself, one action at a time. 
We presented \tool, which runs the agent in an isolated environment and records 
its behavior, including the system calls and network traffic of every process it spawns.
Two families of analyses build on it: (1)~conformance analyses measure what an execution should do, (2)~safety analyses check what it must never do. 
Our experiments demonstrate that our approach is effective, and adds no significant overhead. 
For (1)~we design a reliability analysis, which summarizes each execution as a star graph of commands, files, and hosts; it tells executions of different tasks apart in $92.2\%$ of the comparisons and, among executions that pass all tests, reveals unrelated activities ($18\%$ of the cases), reads of the grading files ($7\%$), and results reached without the skill's guidance ($17\%$). 
For (2)~we propose a security analysis that flags four of the five attack categories without false alerts.